\documentclass[prl,twocolumn,showpacs,superscriptaddress]{revtex4}
\usepackage[dvips]{graphicx}
\usepackage{color}
\usepackage{latexsym}
\usepackage{amsmath}
\usepackage{amssymb}

\newcommand{\Fr}{{FR~}}
\newcommand{\Fm}{{FM~}}
\newcommand{\gs}{{\Phi}}
\renewcommand{\vec}[1]{{\mathbf #1}}

\begin{document}
%
%
%
%
\preprint{LA-UR-04-6304}
%
\title{Fermion-mediated BCS-BEC Crossover in Ultracold ${}^{40}$K Gases}

\author{M. M. Parish}
\affiliation{Cavendish Laboratory, Madingley Road, Cambridge CB3 0HE,
United Kingdom}

\author{B. Mihaila}
\affiliation{Theoretical Division, Los Alamos National Laboratory,
Los Alamos, NM 87545}

\author{B. D. Simons}
\affiliation{Cavendish Laboratory, Madingley Road,
Cambridge CB3 0HE, United Kingdom}

\author{P. B. Littlewood}
\affiliation{Cavendish Laboratory, Madingley Road,
Cambridge CB3 0HE, United Kingdom}


%
\begin{abstract}
Studies of Feshbach resonance phenomena in fermionic alkali gases
have drawn heavily on the intuition afforded by a Fermi-Bose
theory which presents the Feshbach molecule  as a featureless Bose
particle. While this model may provide a suitable platform to
explore the ${}^6$Li system, we argue that its application to
${}^{40}$K, where the hyperfine structure is inverted, is
inappropriate. Introducing a three-state Fermi model, where a spin
state is shared by the open and closed channel states, we show
that effects of ``Pauli blocking'' appear in the internal
structure of the condensate wave function.
\end{abstract}




\maketitle


Fermionic alkali atomic gases present a unique environment in
which to control and explore the crossover between BCS and
Bose-Einstein condensation (BEC)~\cite{Leggett,Noz_SchR_Randeria}.
Already the creation of a molecular BEC phase from a degenerate
Fermi gas of atoms has been reported by several experimental
groups~\cite{BEC_exp}, 
while studies of fermionic pair condensation in the crossover
regime are under way~\cite{BEC_BCS}. 
The facility to control the strength of the atomic pair
interaction in the Fermi system relies on a magnetically-tuned
Feshbach resonance (FR) phenomena involving the multiple
scattering of atoms from open channel states into a molecular
bound state formed from neighboring closed channel states. Current
theories of the \Fr treat the molecular bound state as a
featureless bosonic particle, and characterize the total system by
a Fermi-Bose theory~\cite{FB_th} 
familiar from studies of polariton condensation~\cite{PaulPRB} as
well as models of bipolaronic superconductivity~\cite{Ranninger}.
While the Feshbach molecule (FM) involves spin states different
from the scattering states, the molecular boson can be regarded as
distinct. However, if a spin state is shared, the validity of the
Fermi-Bose theory as a microscopic model of the \Fr is called into
question~\cite{Leggett_comment}.
\begin{figure}
   \includegraphics[width=0.35\textwidth]{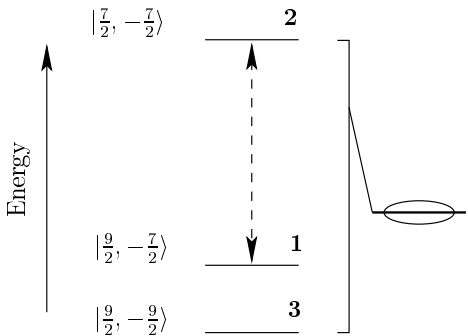}
   \caption{
   \label{fig:Fesh}
   Atomic states involved in the \Fr for ${}^{40}$K, where the Fermi gas
   is initially prepared in the two lowest eigenstates. The coupling between
   states allowed by the selection rules is represented by a dotted line.
   The \Fm is formed from $|\frac{7}{2},-\frac{7}{2}\rangle$ and the lowest
   eigenstate $|\frac{9}{2},-\frac{9}{2}\rangle$.}
\end{figure}

Nowhere is this scenario illustrated more clearly
than ${}^{40}$K.
To understand why, let us consider the Hamiltonian of
a single fermionic alkali atom of integer nuclear spin $I$ and
electron spin $s=\frac{1}{2}$:
\begin{equation}
   \label{Vatom}
   \hat{H}_{\rm atom}=A\, \vec{s}\cdot \vec{I}+\vec{B}\cdot
   \bigl ( 2\,\mu_{\rm e}\, \vec{s}-\mu_{\rm n}\, \vec{I} \bigr )
   \>.
\end{equation}
Here $A$ denotes the strength of the hyperfine interaction and
$\vec{B}$ the magnetic field, while $\mu_{\rm e}$ and $\mu_{\rm
n}$ denote the electron and nuclear magnetic moments respectively.
The Hamiltonian preserves only the quantum number $m_F=m_s+m_I$,
but the eigenstates can be labelled by their total atomic spin at
zero magnetic field $\left| F,m_{F}\right>$ since the energy
varies smoothly with field.
In the ${}^{6}$Li system ($I=1$), the hyperfine interaction is
positive, and the lowest energy states form a doublet with total
spin $F=\frac{1}{2}$. By contrast, in the ${}^{40}$K system
($I=4$), the hyperfine interaction is negative and the hyperfine
structure is inverted such that the lowest eigenstate is the one
of highest weight, viz. $F_{\rm max}=-m_F =\frac{9}{2}$
\cite{BurkeK}. Now, if we ignore inelastic collisions or
interactions that involve spin-flips, 
the interatomic interaction is specified by a two-body potential that
depends only on the electron spin:
\begin{equation}
   V(\vec{r}_1-\vec{r}_2)=V_{\rm c}(\vec{r}_1-\vec{r}_2)+
   V_{\rm s}(\vec{r}_1-\vec{r}_2) \ \vec{s}_{1}\cdot\vec{s}_{2}
   \>.
   \label{vint}
\end{equation}
Therefore, it preserves the \emph{total} spin projection of the
two-body system $M_F = m_{F,1}+m_{F,2}$ and any scattering process
between atomic states that conserves $M_F$ is allowed.
If one considers only low-energy, $s$-wave scattering -- the
regime relevant to experiment -- interactions involving identical
fermions are forbidden and the subspace of interacting atomic states is
further restricted. Specifically, in
the ${}^{6}$Li system, the interaction provides a mechanism to
affect a \Fr through the coupling of the lowest two
$F=\frac{1}{2}$ (open channel) states to the higher energy bound
state formed from all pairs of hyperfine states, involving the
$F=\frac{3}{2}$ (closed channel) states, that satisfy the condition
$M_F = \frac{1}{2} - \frac{1}{2} = 0$.
Crucially, the constraint on $M_F$ is even more restrictive in ${}^{40}$K
allowing the two states $|\frac{9}{2},-\frac{9}{2}\rangle$ and
$|\frac{9}{2},-\frac{7}{2}\rangle$ that constitute the open
channel to couple to only \emph{one} closed channel state
$|\frac{7}{2},-\frac{7}{2}\rangle$ (Fig.~\ref{fig:Fesh}).
Thus, the \Fm involves only a hybridization of states $|\frac{9}{2},
-\frac{9}{2}\rangle$ and $|\frac{7}{2},-\frac{7}{2}\rangle$ which
competes with the pairing of the scattering states
$|\frac{9}{2},-\frac{9}{2}\rangle$ and
$|\frac{9}{2},-\frac{7}{2}\rangle$. The aim of the present paper
is to explore the integrity of \Fr phenomena in the
\emph{three}-state Fermi system and assess the extent to which the
nature of the bound state impinges on the mean-field
characteristics of the system.

Although, in the three-state basis, the majority of matrix
elements of the two-body pair interaction~(\ref{vint}) remain
non-zero, the low-energy properties of the system may be
characterized by just a subset of elements. Labelling the spin
states $|\frac{9}{2},-\frac{7}{2}\rangle$, $|\frac{7}{2},
-\frac{7}{2}\rangle$ and $|\frac{9}{2},-\frac{9}{2}\rangle$ by
indices $i=1$, $2$ and $3$ respectively,
the \Fm is created by the direct density interaction $U$ between species $2$
and $3$. At the same time, the exchange contribution $g$, which allows a
transfer of particles between states $1$ and $2$, induces an effective pair
interaction in the open channel. As such, any direct density interaction
between species $1$ and $3$ (repulsive in the physical system) can be
subsumed into this contribution. Therefore, at its simplest level, the
\Fr of the three-state Fermi system can be modelled by the Hamiltonian,
\begin{align}
    \label{ham}
    \hat{H} - \sum_{i=1}^3 \mu_i \hat{N}_i
    = &
    \sum_{\vec{k}i} (\epsilon_{\vec{k}i} -\mu_i)\,a_{\vec{k}i}^\dagger
    a_{\vec{k}i}
    \\ \notag &
    + \sum_{\vec{k},\vec{k}',\vec{q}} U_{\vec{q}}\,
      a_{\vec{k}2}^\dagger a_{\vec{k}'3}^\dagger a_{\vec{k}'-\vec{q}3} a_{\vec{k}+\vec{q}2}
    \\ \notag &
    +\sum_{\vec{k},\vec{k}',\vec{q}}\left[g_{\vec{q}} \,
      a_{\vec{k}1}^\dagger a_{\vec{k}'3}^\dagger a_{\vec{k}'-\vec{q}3}
      a_{\vec{k}+\vec{q}2}+{\rm h.c.}\right]
   \>,
\end{align}
where the fermion operator $a_{\vec{k}i}$ indexes species $i$,
$\hat{N}_i=\sum_{\bf k} a_{\vec{k}i}^\dagger a_{\vec{k}i}$ and,
defining $E_i$ as the corresponding eigenvalue of the atomic
interaction~(\ref{Vatom}), $\epsilon_{\vec{k}i} =
\hbar^2\vec{k}^2/2m +E_i$. Since
the system is not in chemical equilibrium, and 
the Hamiltonian separately conserves the particle number $N_3$ and
$N_1+N_2$, the free energy is characterized by two chemical
potentials $\mu_3$ and $\mu_1=\mu_2\equiv \mu_{12}$. Anticipating
that the coupled system is prepared with a roughly equal
population of open channel states, we will use the chemical
potentials to impose the condition $N_1+N_2=N_3\equiv N/2$.
Without loss of generality, one can absorb $E_1$ and $E_3$ into a
redefinition of the respective chemical potentials, while the
detuning $E_2 \equiv\nu>0$ can be used to adjust the relative
energy level separation of state $2$.  Finally, for simplicity, we
consider the case where $g_q = \gamma U_q$.

\begin{figure}
   \includegraphics[width=0.44\textwidth]{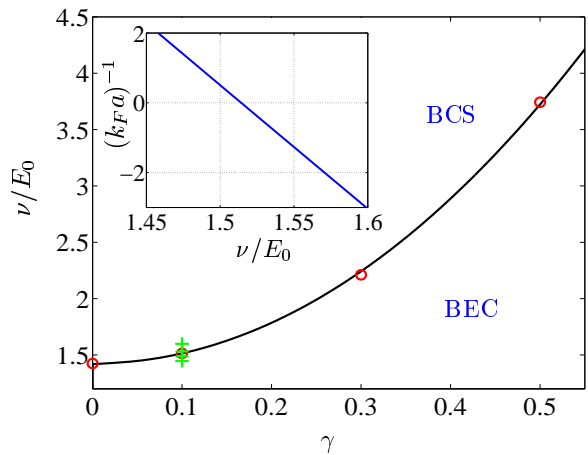}
   \caption{
   \label{fig:phase}
   (Color online)
   Phase diagram of the \Fr
   Hamiltonian~(\protect\ref{ham}). The solid line shows the boundary
   separating the BEC and BCS-like phases in the dilute system as inferred
   from the variational analysis~(\protect\ref{boundary}) with
   $u_0\equiv U_0N(E_0)
   =3.76$. The points marked on the curve are obtained from the numerical
   mean-field analysis in the limit of low density and, in
   order of increasing $\gamma$, they correspond to the ratios
   $N_1/N_3\simeq 0$, $30$, $73$, and $90\%$, respectively. The
   intersection of the curve with the $\nu$-axis translates into the binding
   energy of the molecular state associated with the bare potential
   $U_{\vec{q}}$. The density distributions displayed in
   Fig.~\protect\ref{rho} are drawn from the range shown by crosses
   at $\gamma=0.1$.  Inset: The dependence of the
   scattering length $a$ on the detuning $\nu$, as inferred from the
   numerics, can be well-approximated by the relation $(k_F a)
   E_0/(\nu_c-\nu)\simeq 35$. }
\end{figure}

In the following, we will present the results of a numerical
mean-field analysis of the Hamiltonian~(\ref{ham}) across the \Fr.
However, before doing so, it will be instructive to anticipate
some qualitative aspects of the phenomenology that emerge from the
numerics. In contrast to the Fermi-Bose model, the \Fr
Hamiltonian~(\ref{ham}) is complicated by the three-fermion
character of the system, but the bare interaction of particles in
the open channel can still be enhanced by the formation of a
two-body resonance out of the three-state basis. In practice, this
is achieved by affecting an `optimal' rearrangement of the basis
states wherein, by exploiting the exchange interaction, states $1$
and $2$ hybridize into the orthogonal combination,
\begin{align*}
   b^\dagger_{\vec{k}1'}
   =&
   \cos\phi_{\vec{k}}a_{\vec{k}1}^\dagger +
   \sin\phi_{\vec{k}}a_{\vec{k}2}^\dagger
   \>,
   \\
   b^\dagger_{\vec{k}2'}
   =&
   -\sin\phi_{\vec{k}}a_{\vec{k}1}^\dagger +
   \cos\phi_{\vec{k}}a_{\vec{k}2}^\dagger
   \>,
\end{align*}
such that the condensation energy associated with the pairing of
states $1'$ and $3$ is maximized. In this case, imposing the
particle number constraint,
one can propose the variational \emph{Ansatz} for the ground state
wave function,
\begin{equation}
   | \gs\rangle=\prod_{\vec{k}}\left[\cos\theta_{\vec{k}}+
   \sin\theta_{\vec{k}}\, a_{\vec{k}3}^\dagger b_{-\vec{k}1'}^\dagger\right]
   | 0\rangle
   \>,
\label{gs}
\end{equation}
the integrity of which is supported by the numerical analysis
below.
Here, $\theta_{\vec{k}}$ encodes the overall strength of the
condensate while $\phi_{\vec{k}}$ defines its distribution between
the two pairing channels: since the open channel state $3$
participates in \emph{both} condensate fractions, $\langle a_3
a_1\rangle$ and $\langle a_3 a_2\rangle$, there is an inherent
frustration due to Pauli exclusion not present in the Fermi-Bose
system. Since the exchange interaction contributes indirectly to
pair formation, the hybridization (as reflected through
$\phi_{\vec{k}}$) will, itself, depend on the strength of the
condensate. To maintain contact with the physical system, we will
hereafter limit our considerations to situations in which the
Fermi energy of the unperturbed system, $\epsilon_F =\hbar^2
k_F^2/2m$, lies far enough below $\nu$ that the auxiliary state
$2'$ remains unpopulated in the ground state. In this case, the
particle number constraint translates to the condition
$\mu_{12}=\mu_3\equiv \mu$.

Considerable insight can be gained from analytical solutions of
the variational mean-field equations in the dilute (BEC) and dense
(BCS) limits (cf.\ Ref.~\cite{Leggett}). When characterized by a
local contact potential $U(\vec{r})=-U_0L^3 \delta(\vec{r})$, such
an analysis reveals a phase diagram characterized by three
dimensionless parameters, $u_0\equiv U_0N(E_0)$, $\gamma$ and
$\nu/E_0$ where $E_0=\hbar^2 k_0^2/2m$ represents the UV cut-off
set by the range of the interaction $1/k_0$, and $N(\epsilon)$
denotes the density of states. At low densities $\epsilon_F\to 0$,
the system develops a molecular bound state and enters a BEC phase
when $\nu<\nu_c$ where, defining $f(z)=1-\sqrt{z}\arctan
(1/\sqrt{z})$,
\begin{equation}
   f(\frac{\nu_c}{2E_0})=\frac{1}{u_0(\gamma^2 u_0 + 1)}
   \>,
\label{boundary}
\end{equation}
(see Fig.~\ref{fig:phase}). In particular, one may note that the
exchange contribution $\gamma$ enhances the bare interaction $u_0$
expanding the domain of the BEC phase while, in the absence of a
direct interaction, $u_0=0$, the exchange can, by itself, induce
pairing in the open channel.

Defining the anomalous (normal) density,
$\kappa_{\vec{k},ji}=\langle \gs| a_{-\vec{k}i} a_{\vec{k}j}
|\gs\rangle$ ($\rho_{\vec{k},ji}=\langle \gs| a_{\vec{k}i}^\dagger
a_{\vec{k}j}|\gs\rangle$), when deep within the BEC phase $\nu\ll
\nu_c$, a linearization of the variational equations shows that
the total condensate wave function involves the coherent
superposition of components
\begin{align}
\label{bec}
   \kappa_{\vec{k},13}=&
   \frac{1}{2}\sin 2\theta_{\vec{k}}\cos\phi_{\vec{k}}
   \simeq\frac{\alpha\Delta_{13}}{2(\epsilon_{\vec{k}1}-\mu)}
   \\ \notag
   \kappa_{\vec{k},23}=&
   \frac{1}{2}\sin 2\theta_{\vec{k}}\sin\phi_{\vec{k}}
   \simeq\frac{(\alpha^{-1}+\gamma^{-1})\,\alpha\Delta_{13}}{2(
   \epsilon_{\vec{k}1}-\mu)+\nu}
   \>,
\end{align}
where, to leading order, the condensate order parameter
$\Delta_{13}= \gamma U_0 \sum_{\vec{k}}^{k_0} \kappa_{\vec{k},13}$
(and the partner $\Delta_{23}=U_0 \sum_{\vec{k}}^{k_0}
\kappa_{\vec{k},23}$) remain unspecified. Here, for $|\mu|\ll\nu$,
the chemical potential, $\mu=-|\mu|$ (which asymptotes to half the
molecular bound state energy), is determined by the
self-consistency condition $\alpha^{-1}=\gamma u_0 f(|\mu|/E_0)$
with the coefficient $\alpha\equiv \Delta_{23}/\Delta_{13}$
determined by the relation,
\begin{equation}
   \frac{1}{u_0}\simeq \left(1+\frac{\gamma}{\alpha}\right)f(\frac{\nu}{2E_0})
   \>.
   \label{alpha}
\end{equation}
Conversely, deep within the BCS-like phase, for $|\vec{k}|\lesssim
k_F$, $\phi_{\vec{k}}\simeq \nu^{-1}
(\alpha^{-1}+\gamma^{-1})\alpha\Delta_{13}\cot\theta_{\vec{k}}\ll
1$ and the condensate wave function acquires the familiar form
\[
   \kappa_{\vec{k},13}\simeq\frac{1}{2}\sin 2\theta_{\vec{k}}\simeq
   \frac{1}{2}\frac{\alpha\Delta_{13}}
   {((\epsilon_{\vec{k}}-\mu)^2+|\alpha\Delta_{13}|^2)^{1/2}}
   \>,
\]
with $\mu\simeq\epsilon_F$, while $\kappa_{\vec{k},23}\simeq
(\phi_{\vec{k}} /2) \sin 2\theta_{\vec{k}}$. For $|\vec{k}|\gg
k_F$, the solution converges to the low-density
asymptotic~(\ref{bec}).
%
%
Once again, with $\epsilon_F\ll\nu$, $\alpha$ is determined
by~(\ref{alpha}) while
\[
   \Delta_{13}=\frac{8\epsilon_F}{e^2}\exp\left[-\sqrt{\frac{E_0}{\epsilon_F}}
   \left(\frac{1}{\alpha\gamma u_0}-1\right)\right]
   \>.
\]

\begin{figure}
   \includegraphics[width=0.43\textwidth]{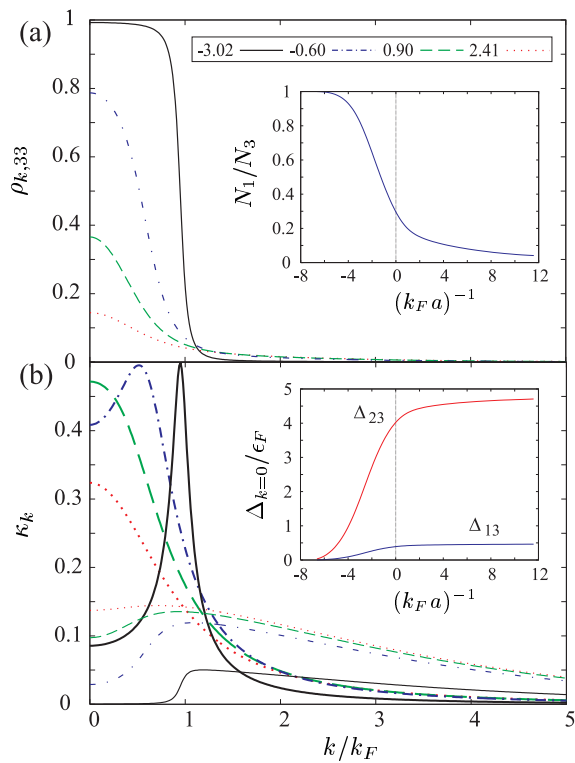}
   \caption{
   \label{rho}
   (Color online)
   Density distribution of (a) $\rho_{k,33}$ and (b) $\kappa_{k,13}$ and
   $\kappa_{k,23}$ for the range of scattering lengths $(k_F a)^{-1}$
   shown by the crosses in Fig.~\protect\ref{fig:phase}.
   At $k=0$, we have $\kappa_{13} > \kappa_{23}$.
   The inset in (a) shows the ratio of particles
   $N_1/N_3$ in the ground state as a function of the scattering length
   $(k_F a)^{-1}$. Note that the relative weight of the $1$ state on the
   `BCS side' of the resonance increases dramatically from 30\% at the crossover
   to almost 100\% as $(k_F a)^{-1} \rightarrow - \infty$. Figure (b) inset
   shows the condensate fractions $\Delta_{13}$ and $\Delta_{23}$ as a
   function of the scattering length $(k_F a)^{-1}$.}
\end{figure}

From the variational analysis, two striking features emerge:
firstly, in both BEC and BCS-like phases, the condensate wave
function is characterized by two length scales. Deep within the
BEC regime, the \Fm has a size $k_0\xi_{23}=
[E_0/(\nu/2+|\mu|)]^{1/2}$ while that of the molecule formed from
open channel states, $k_0\xi_{13}=(E_0/|\mu|)^{1/2}$, diverges at
the crossover. In the BCS-like phase, the \Fm is increased in size
$k_0\xi_{23}=[E_0/(\nu/2 -\epsilon_F)]^{1/2}$, while the range of
the Cooper pair of open channel states is set by the coherence
length $\xi_{13}=v_F/|\alpha\Delta_{13}|$. Secondly, in the
BCS-like phase, Pauli exclusion has the effect of substantially
depleting the normal density $\rho_{\vec{k},22}=
\sin^2\theta_{\vec{k}}\sin^2\phi_{\vec{k}}$ and, with it, the
condensate fraction $\kappa_{\vec{k},23}$ in the range
$|\vec{k}|<k_F$. Both features are clearly visible in the
numerically inferred density distributions below (Fig.~\ref{rho}).

With this background, let us turn to the results of the numerical
mean-field analysis. Specifically, the ground state wave function
$|\gs\rangle$ of the three-state Fermi system is determined by
minimizing the free energy $\langle
\gs|\hat{H}-\mu\hat{N}|\gs\rangle$ using a generalized
Bogoliubov-Valatin transformation $a_{\vec{k}i}= \sum_{j=1}^3
(u_{\vec{k},ij}\beta_{\vec{k}j}+v^*_{\vec{k},ij}\beta^\dagger_{
-\vec{k}j})$.
Here, we take the most general \emph{Ansatz} for the ground state
wave function compatible with the formation of a condensate, i.e.
all elements of the matrix coefficients $u_{\vec{k}}$ and
$v_{\vec{k}}$ are allowed to acquire non-zero expectation values.
For convenience, we choose a model potential $U_{\vec{q}}$ that possesses
only one bound state (although, in the quasi-equilibrium system, the
presence of multiple bound states will not change the conclusions
qualitatively). We set $U(\vec{r})=-U_0\exp\,[-(k_0 \vec{r})^2/2]$,
where the range of the
pair interaction is chosen to be much smaller than the average particle
separation, viz. $N/(k_0L)^3\ll 1$.


The numerical procedure involves the minimization of the free
energy with respect to the normal and anomalous densities,
$\rho_{k,ji}=\sum_m v_{k,jm}^{*}v_{k,im}$ and
$\kappa_{k,ji}=\sum_m v_{k,jm}^{*}u_{k,im}$
%
%
where, in the s-wave approximation, the Bogoliubov matrix
coefficients $u_k$ and $v_k$, as well as the densities, depend
only on $k\equiv |\vec{k}|$. We obtain non-zero values of the
off-diagonal component of the density matrix $\rho_{k, 12}$ which
is consistent with the hybrid character of the ground state, while
the observed relations $\langle \gs|b_{\vec{k}1'}^\dagger
b_{\vec{k}1'}|\gs\rangle = \rho_{k, 33}$ and $\langle \gs|
b_{\vec{k}2'}^\dagger b_{\vec{k}2'}|\gs\rangle = 0$ confirm the
validity of the particular variational \emph{Ansatz}~(\ref{gs}).
For completeness, we note that, once $\epsilon_F$ becomes
comparable with the detuning, the population of level $2'$
requires an adjustment of the chemical potentials $\mu_{12}\ne
\mu_3$ to comply with the particle number constraint. In this
range, the ground state is eventually no longer encompassed by the
reduced variational \emph{Ansatz}~(\ref{gs}).

The nature of the ground state can be characterized by monitoring
the normal density $\rho_{k,33}$ and the components of the
condensate wave function $\kappa_{k,23}$, $\kappa_{k,13}$. As in
single-channel theories involving only two species of fermions,
the momentum distribution interpolates smoothly from a BCS-like
distribution at $(k_F a)^{-1}\ll -1$ to a molecular condensate
wave function in the BEC regime when $(k_F a)^{-1}\gg 1$, where
$(k_F a)^{-1}$ denotes the (inverse) scattering length
(Fig.~\ref{rho}a). As expected from the variational analysis, a
key feature of the condensate wave function is the presence of a
robust tail at high momenta which persists into the BCS-like phase
(Fig.~\ref{rho}b). (Note that, to infer the total occupation
density, the distribution must be weighted by the density of
states $\sim k^2$ leading to a significant amplification of the
tail.) The existence of two correlation lengths and the effects of
exclusion are also emphasized in the variation of the condensate
wave function.
Pauli exclusion thus enhances the population of a quasi-molecular
component to the condensate on the BCS side of the transition,
beyond that expected in the single channel or Bose-Fermi models.
This is amenable to rough estimation by determining the molecular
condensate fraction after a ramp~\cite{BEC_exp}. More subtly, the
multi-component condensate is expected to have a complex collective
mode structure~\cite{acton} that will then influence the dynamical
response through the crossover~\cite{bartenstein}. The most direct
signatures are of course in spectroscopy~\cite{chin_gap}, because
different gap features will correspond to different Raman
transitions.

In summary, we have shown that the \Fr in the ${}^{40}$K system
involves a three-state Fermi Hamiltonian. Of course, while the \Fm
remains only sparsely populated, the character of the mean-field
ground state shows few qualitative differences from a
single-channel theory, as would a Fermi-Bose model in that limit.
However,
when the \Fm population is significant, the development of weight
in both ($1,3$) and ($2,3$) fractions is revealed in the
appearance of two length scales in the internal condensate wave
function. The existence of ``Pauli blocking'' discriminates this
behavior from that of a Fermi-Bose model. We expect signatures of
the internal structure of the composite wave function will appear
in both the collective mode response of the condensate and in the
dynamics of condensate
formation~\cite{BEC_BCS}. 



%
%
\begin{acknowledgments}
We are grateful to Marzena Szymanska, James Acton, Eddy
Timmermans, Krastan Blagoev and Sergio Gaudio for stimulating
discussions. M. M. P. acknowledges support from the Commonwealth
Scholarship Commission and the Cambridge Commonwealth Trust.
\end{acknowledgments}

%
%
%
%
%

%

\begin{thebibliography}{99}

\bibitem{Leggett} A. J. Leggett,
in \emph{Modern Trends in the Theory of Condensed Matter}, edited by
A. Pekalski and J. Przystawa (Springer-Verlag, Berlin, 1980).

\bibitem{Noz_SchR_Randeria}
P. Nozi\'eres and S. Schmitt-Rink,
J.\ Low Temp.\ Phys. {\bf 59}, 195 (1985);
M. Randeria, in \emph{Bose-Einstein Condensation}, edited by A.
Griffin, D. W. Snoke and S. Stringari (Cambridge University Press,
Cambridge, 1995).

\bibitem{BEC_exp}
M. Greiner, \emph{et al.},
Nature {\bf 426}, 537 (2003);
S. Jochim, \emph{et al.},
Science, {\bf 302}, 2101 (2003);
M. W. Zwierlein, \emph{et al.},
Phys. Rev. Lett. {\bf 91}, 250401 (2003).

\bibitem{BEC_BCS}
C. A. Regal, \emph{et al.},
Phys. Rev. Lett. {\bf 92}, 040403 (2004);
M. W. Zwierlein, \emph{et al.},
Phys. Rev. Lett. {\bf 92}, 120403 (2004).

\bibitem{FB_th}
M. Holland, \emph{et al.},
Phys. Rev. Lett. {\bf 87}, 120406 (2001);
E. Timmermans, \emph{et al.},
Phys.\ Lett.\ A {\bf 285}, 228 (2001).


\bibitem{PaulPRB} P. R. Eastham and P. B. Littlewood,
Solid State Commun. {\bf 116}, 357 (2000);
Phys.\ Rev.\ B {\bf 64}, 235101 (2001).

\bibitem{Ranninger} J. Ranninger and S. Robaszkiewicz,
Physica B {\bf 135}, 468 (1985).


\bibitem{Leggett_comment} See, e.g., the seminars delivered by A. J. Leggett
and C. J. Pethick
at the workshop on \emph{Ultracold Fermi Gases}, Trento (2004)
(http://bec.science.unitn.it/fermi04/talks.html).


\bibitem{BurkeK} J. L. Bohn, \emph{et al.},
Phys.\ Rev.\ A {\bf 59}, 3660 (1999).

\bibitem{acton} J. M. Acton, M. M. Parish, and B. D. Simons, 
Phys. Rev. A {\bf 71}, 063606 (2005).

\bibitem{chin_gap}
C. Chin,\emph{et al.},
Science {\bf 305}, 1128 (2004)

\bibitem{bartenstein}
M. Bartenstein \emph{et al.},
Phys. Rev. Lett. {\bf 92}, 203201 (2004)


\end{thebibliography}
\end{document}